\documentclass{elsart}
\usepackage{natbib}

\usepackage[dvips]{graphics}

\newcommand{\beq}{\begin{equation}}     \newcommand{\eeq}{\end{equation}}
\newcommand{\beqa}{\begin{eqnarray}}    \newcommand{\eeqa}{\end{eqnarray}}
\newcommand{\bde}{\begin{description}}  \newcommand{\ede}{\end{description}}
\newcommand{\ben}{\begin{enumerate}}    \newcommand{\een}{\end{enumerate}}

\newcommand{\ltsim}{\mbox{${}^{<}_\sim$}}
\newcommand{\kT}{\mbox{$k_{\rm B}T$}}

\begin{document}
\begin{frontmatter}
\title{Autonomous free-energy transducer \\
working under thermal fluctuations}
\author{Ken Sekimoto}
\address{Groupe Physico-Chimie Th\'eorique, ESPCI\\
10 rue Vauquelin, F75231 Paris Cedex 05, 
France\thanksref{a}}
\ead{sekimoto@turner.pct.espci.fr}
\thanks[a]{On leave from Universit\'e Louis Pasteur, Strasbourg, France.}
\begin{abstract}
By a modular combination of mesoscopic detectors and gates, 
we present a thoughtful pump 
that transports gas particles against the difference of 
their density at the expense of the diffusion of another 
species of gas particles.
We also discuss briefly the relevance of the model to the
study of structure-function relationship of the
biomolecular machines.
\end{abstract}

\end{frontmatter}
%%%%%%%%%%%%%%%%%%%%%%%%%%%%%%%%%%%%%%%%%%%%%%%%%%%%%%%%%%%%%%%%%%%%%%%%%%

%%%%%%%%%%%%%%%%%%%%%%%%%%%%%%%%%%%%%%%%%%%%%%%%%%%%%%%%%%%%%%%%%%%%%%%%%%%%%%%
\section{Introduction}
%%%%%%%%%%%%%%%%%%%%%%%%%%%%%%%%%%%%%%%%%%%%%%%%%%%%%%%%%%%%%%%%%%%%%%%%%%%%%%%
\label{sec:introduction}

Biological molecular machines such as ion pumps, molecular motors (myosin,
F$_1$ motor, etc.) or signal transducers (G-proteins) have attracted 
a lot of interest among physicists as well as biologists  
\citep{cell}.
The molecular machines works under thermal 
fluctuations, where the latter
serve as an energy source \citep{sekimoto97}  
for the free energy transduction \citep{eisenberg85}\footnote{
Upon a single ATP hydrolysis reaction {\it in vivo}, 
about 70\% of the maximally available work (20\kT with 
\kT$\simeq$ 4pN$\cdot$nm) 
is ascribed to the change of the mixing entropy.}
as well as for the thermal activation \citep{Wong}. 
Moreover, the cycle of free-energy transduction in 
a single molecular machine requires no external control.
As many of these molecular machines share highly conserved 
three-dimensional structural modules,
it is widely believed that they have been evolved 
from a common ancestors \citep{ancestor}.
This fact suggests that such molecular machines may be 
composed of a few modular structures each of which bears 
some elementary functional role. 
With recent rapid development of nano-biology
and the structural biology, 
the biologists have begun to look for such modular 
structure-function relationships
within a single molecular motors 
\citep{higuchi02,KIF1A04,strain-detector-a}.

From the theoretical physicists side, there is no surprise 
about the mere fact that a molecular machine
without any special forward-backward symmetry
can move in one direction under non-equilibrium 
conditions (e.g. under the ATP hydrolysis),
as stated in the Curie's principle \citep{curie}\footnote{
{\it ``Lorsque certaines causes produisent certains effets, 
 les \'el\'ements de sym\'etrie
 des causes doivent se retrouver dans les effets produits.''}}. 
With this in mind, the question of our interest is 
how we can construct,
in combining well-defined simple functional modules,
an autonomously controlled 
system that can work under thermal fluctuations.
Since the early works by \citet{feynman66} and \citet{buttiker87}, 
several models of autonomous free energy 
transducer have been proposed, and their improvements
through the incorporation of a ``gate'' have been proposed 
\citep{sakaguchi,derenyi99} (see also the review by \citet{reimann}).
These models were, however, rather {\it ad hoc} and the 
modular nature of the construction was not clearly visible.
In the present paper, 
we would construct a free energy transducer 
with being more conscious about the modules and 
taking the object oriented approach \citep{preprint}. 
We start by introducing the concept of 
{``semi-detectors''}, the detectors 
that can perceive a certain external state, but not all, with 
arbitrary sureness despite the thermal fluctuation 
(\S~\ref{sec:detector}). 
We then combine these semi-detectors together with 
the gates, under the designing concept 
that we call the {``bidirectional control''},
to realize an autonomous 
particle transporter (\S~\ref{sec:transducer}). 
Because of the limited space of this special issue, 
the extensive discussion on the possible 
relations to the real biological molecular machines
will be given elsewhere.

%%%%%%%%%%%%%%%%%%%%%%%%%%%%%%%%%%%%%%%%%%%%%%%%%%%%%%%%%%%%%%%%%%%%%%%
\section{semi-detectors in the fluctuating world}
\label{sec:detector} 
%%%%%%%%%%%%%%%%%%%%%%%%%%%%%%%%%%%%%%%%%%%%%%%%%%%%%%%%%%%%%%%%%%%%%%%

\begin{figure}[ht]  
\hspace*{1cm}  
\resizebox{12.00cm}{3.00cm}{\includegraphics*{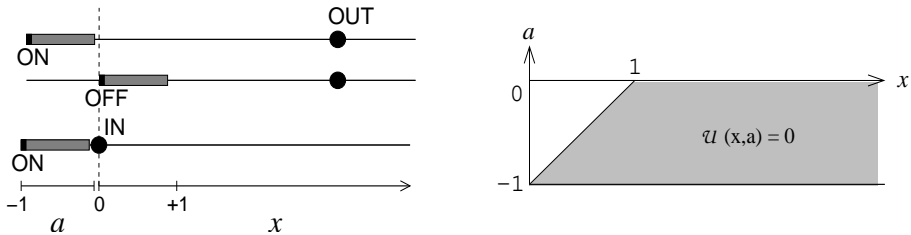}}
\caption{
{\it Left:}
Schematic representation of the semi-detector of absence in
a continuum representation.
The movable object (thick bar) and the ligand particle (thick dot)
are sterically excluding with each other.
$a$ represents the coordinate of the left extremity of the object.
$x=0$ is the detection cite. 
{\it Right:}
The shadowed region on the $(x,a)$-plane 
indicates the accessible phase space, where 
$U(x,a)$ =0. } 
\label{38up23.eps} 
\end{figure}

The detector is the mechanism that correlates the states of the 
outside to those of the inside of the system in question. 
We focus on the case where the detection site admits a single 
ligand particle coming randomly from the outside.
We will limit ourselves to the energetically ``neutral'' detections, 
involving no change of the total energy.
Let us define the mapping from a
state of the outside, $x$, to a subset of the states of the inside, 
$\Phi(x)$.
In the discrete representation,
the states of outside consists of {\sf IN} (i.e. a ligand
particle is on the detection site) and {\sf OUT} (otherwise),
while those states inside consist of {\sf ON} and {\sf OFF}.
The perfect detector would establish the mapping,  
$\Phi({\sf IN})=\{ {\sf ON}\}$ and $\Phi({\sf OUT})=\{ {\sf OFF}\}$.
But these correspondences are too stringent to realize under 
thermal fluctuations.
We will show below that there can be the physical
mechanisms which assure either one of the above correspondences.
We will call the mappings corresponding to such mechanisms 
the semi-detectors.
{\it Semi-detector of absence:} This module functions as the mapping 
$\Phi_{\rm ab}(x)$ that prohibits only the output of
{\sf OFF} under the input of {\sf IN}. That is,
\beq \label{eq:absence{detector}}
\Phi_{\rm ab}(\mbox{{\sf IN}})=\{\mbox{{\sf ON}}\}, \quad
\Phi_{\rm ab}(\mbox{{\sf OUT}})=\{\mbox{{\sf ON}},\mbox{{\sf OFF}}\}.
\eeq
Then the {\sf OFF} surely indicates the 
absence of the particle ({\sf OUT}). 
Such aspect is 
useful for repressive processes like the suppression of DNA
transcription by an repressor \citep{genes8}.
In the continuum representation, one may define $x$ as the position of 
a ligand particle in a half space $0\!\le \!x \!< \!\infty $, where $x=0$ 
corresponds to the detection site (see Fig.~\ref{38up23.eps} {\it Left}).
The state of the inside, $a$, is assumed to be bounded, say,
in the region of $ -1\! \leq\! a\! \leq \! 0$ without loosing generality.
The semi-detector of absence may be realized 
by the steric repulsion between the ligand 
particle and a movable object 
(thick bars in Fig.~\ref{38up23.eps} {\it Left}).
The total energy of the ligand-detector system, $U(x,a)$,
is written as $U(x,a)=0$ for $x\! \ge a \!+\! 1$ and $U(x,a)=\infty$ otherwise.
{\sf OFF} and {\sf ON} states correspond, respectively, to 
$a\! \ge\!  -1\!+\!\delta$ and $-1\! \le\!  a 
\! <\!  -1\!+\!\delta$ with a small $\delta\!  >\! 0$.
% 
%%%%%%%%%%%%%%%%%%%%%%%%%%%
%%%%%%%%%%%%%%%%%%%%%%%%%%%

\begin{figure}[ht]  
\hspace*{1cm}  
\resizebox{12.00cm}{3.00cm}{\includegraphics*{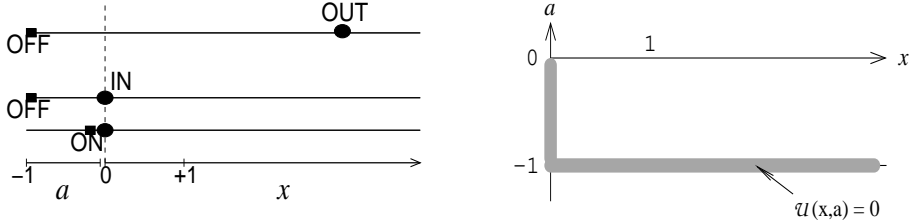}}
\caption{{\it Left:} 
Schematic representation of the semi-detector of presence.
$a$ and $x$ represent, respectively, the position of
the movable object (filled square) and that of the ligand particle 
(thick dot).
Unless the particle is within 
the proximity of the detection cite ($0\le x<\delta$) with 
small $\delta$, the movable object is constrained at $a\simeq -1$
even under thermal agitations.
{\it Right:}
The accessible phase space 
is indicated as a shadowed region, where
$U(x,a)\sim \kT$. 
} 
\label{38low24.eps} 
\end{figure}

%
{\it Semi-detector of presence:} This module functions complementarily to 
the former semi-detector. The mapping 
$\Phi_{\rm pr}(x)$ that it defines prohibits only the output of
{\sf ON} under the input of {\sf OUT}. That is,

\beq \label{eq:presence{detector}}
\Phi_{\rm pr}(\mbox{{\sf IN}})=\{\mbox{{\sf ON}},\mbox{{\sf OFF}}\}, \quad
\Phi_{\rm pr}(\mbox{{\sf OUT}})=\{\mbox{{\sf OFF}}\}.
\eeq
Then the {\sf ON} surely indicates the 
presence of the particle ({\sf IN}). 
Such aspect is useful for active processes like the 
uptake of ATP by a molecular motor.
In the continuum representation, 
the semi-detector of presence may be realized by the compensation
of a strong restoring potential for the movable object, 
 $M\phi(a)$ ($M\!\gg\! \kT$), by the 
strong attractive interaction energy, $-M\phi(a-x)$, 
between the ligand particle and
the movable object (see Fig.~\ref{38low24.eps} {\it Left}). 
Here $\phi(z)$ is defined on $-\infty< \! z\!\le\! 0$ so that
$\phi(z)=0$ for $z\!\le\! -1$ and that $\phi(z)$ increases 
monotonically from $\phi(-1)=0$ to $\phi(0)=1$.
%
%%%%%%%%%%%%%%%%%%%%%%%%%%%
%%%%%%%%%%%%%%%%%%%%%%%%%%%
Energetically, 
this movable object can take the value $a \ge -1+\delta$
only when the particle is near the binding site,
that is $0\le x\le \delta$ with a small $\delta>0$ so that 
$M \phi(\delta) \ltsim \kT$\footnote{One can verify this by drawing
$M\{\phi(a)-\phi(a-x)\}$ {\it vs} $a$ for various values
of $x$.}.
Therefore {\sf ON} and {\sf OFF} correspond, respectively, to 
$a\ge -1+\delta$ and $-1\le a < -1+\delta$.
The compensation mechanism similar to the one 
discussed here has been discussed and called 
{\it induced fit} 
by \citet{koshland73} in the context of the 
ligand binding by enzymes.

The above two semi-detectors establishes the 
 correlations between the outer world of the detectors 
and the detectors themselves in the way that the semi-detectors
 represent certain error-free information despite the thermal
fluctuations.
If the state variables of two semi-detectors of presence, 
say $a_1$ and $a_2$, respectively, are coupled energetically
with each other, 
the resulting system might function to realize the 
{``cooperative 
binding''} \citep{monod65,koshland66} or 
the  {``exchange of binding''} \citep{eisenberg85}, depending on
the nature of the coupling.
%

%%%%%%%%%%%%%%%%%%%%%%%%%%%%%%%%%%%%%%%%%%%%%%%%%%%%%%%%%%%%%%%%%%%%%%%
\section{Free-energy transducer based on the {semi-detector}s of
presence and the gates}
%%%%%%%%%%%%%%%%%%%%%%%%%%%%%%%%%%%%%%%%%%%%%%%%%%%%%%%%%%%%%%%%%%%%%%%
\label{sec:transducer}

\begin{figure}[ht]  
\hspace*{2.5cm}  
\resizebox{7.00cm}{4.50cm}{\includegraphics*{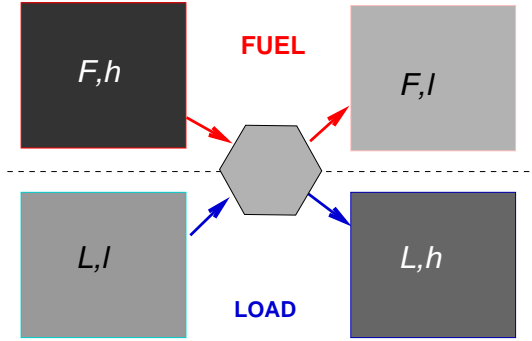}}
\caption{Schematic diagram of how the free-energy transducer 
(the hexagon) 
pumps up the {\sf load} particles at the expense of 
the diffusion of the {\sf fuel} particles. See the text.}
\label{16.eps} 
\end{figure}

In order to focus on the subject of the autonomous control, 
we would like to avoide energetic aspects as far as possible.
To this end, we will construct a pump of the ideal ``{\sf load}''(L) 
gas particles from a dilute reservoir, $({\rm L},{\rm l})$,
to a dense reservoir, $({\rm L},{\rm h})$, 
at the expense of the diffusion of the 
ideal ``{\sf fuel}''(F) particles
from a dense reservoir, $({\rm F},{\rm h})$,
to a dilute reservoir, $({\rm F},{\rm l})$, 
see Fig~\ref{16.eps}.
%%%%%%%%%%%%%%%%%%%%%%%%%%%
%%%%%%%%%%%%%%%%%%%%%%%%%%%
If we denote the chemical potentials of these reservoirs by
$\mu_{\rm L,{\rm l}}$, $\mu_{\rm L,h}$, $\mu_{\rm F,h}$ and  
$\mu_{\rm F,{\rm l}}$, respectively, 
then the decrease of the total Gibbs' free energy
in the course of the transport of 
$\Delta N_{L,{\rm l}\to{\rm h}}$ of the {\sf load} particles 
at the expense of $\Delta N_{F,{\rm h}\to{\rm l}}$ 
of the {\sf fuel} particles is 
$(\mu_{\rm F,h}-\mu_{\rm F,{\rm l}})\Delta N_{F,{\rm h}\to{\rm l}}- $
$ (\mu_{\rm L,h}-\mu_{\rm L,{\rm l}}) \Delta N_{L,{\rm l}\to{\rm h}}$. 
Below we will present a completely symmetric model 
in which the roles of the {\sf fuel} and the {\sf load} are 
totally exchangeable depending on the relative magnitude of 
$\mu_{\rm F,h}-\mu_{\rm F,{\rm l}}$ and 
$\mu_{\rm L,h}-\mu_{\rm L,{\rm l}}$\footnote{Under this symmetry,
the currents of the active transport of the {\sf load} particles, 
$J_{\rm L, {\rm l}\to{\rm h}}$, and that of the passive diffusion of 
the {\sf fuel} particles, $J_{\rm F, {\rm h}\to{\rm l}}$, 
are given in the linear non-equilibrium thermodynamics as 
$J_{\rm L,{\rm l}\to{\rm h}}= $
$L_{\rm asy}\Delta \mu_{\rm asy}- L_{\rm sym}\Delta \mu_{\rm sym}$
and 
$J_{\rm F,{\rm h}\to{\rm l}}=$ 
$L_{\rm asy}\Delta \mu_{\rm asy}+ L_{\rm sym}\Delta \mu_{\rm sym}.$
Here 
$L_{\rm asy}\ge 0$ and $L_{\rm sym}\ge 0$ 
are the kinetic coefficients, and $\Delta \mu_{\rm asy}\equiv$ 
$\mu_{\rm F,h}-\mu_{\rm F,{\rm l}}-\mu_{\rm L,h}+\mu_{\rm L,{\rm l}}$
and $\Delta \mu_{\rm sym}\equiv$ 
$\mu_{\rm F,h}-\mu_{\rm F,{\rm l}}+\mu_{\rm L,h}-\mu_{\rm L,{\rm l}}$. 
This pump can work almost reversibly near the stalled condition, 
$\Delta \mu_{\rm asy}=0$, if $L_{\rm sym}$ is sufficiently small.}.

\begin{figure}[ht]  
\hspace*{2.5cm}  
\resizebox{8.80cm}{4.00cm}{\includegraphics*{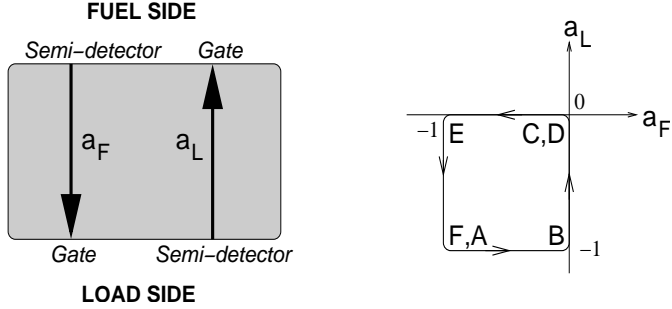}}
\caption{{\it Left:} Scheme of the bidirectional control. 
{\it Right:} Typical cycle realized by the
two independent internal state variables. 
{\sf A}-{\sf F} correspond to those in Fig.~\ref{19new.eps}.}
\label{27.eps} 
\end{figure}

The {\it cyclic} process in the above mentioned pump 
requires at least two internal degrees of 
freedom\footnote{This statement does not contradicts 
with the fact that the existing models 
of heat engine \citep{feynman66,buttiker87} have assumed
only one rotational degree of freedom to lift a load, since such
degree of freedom corresponds to two bounded degrees of freedom.}.
To construct such pump with 
using the semi-detectors of presence as constituent modules, 
the most natural way would be to 
put one semi-detector on the {\sf fuel} side 
(with the variable $a_{\rm F}$) 
and the another one on the {\sf load} side
(with the variable $a_{\rm L})$,  
and to have them to control the access of the particles 
on the opposite sides,
via allosteric couplings \citep{monod65}, see Fig.~\ref{27.eps}.
%
%%%%%%%%%%%%%%%%%%%%%%%%%%%
%%%%%%%%%%%%%%%%%%%%%%%%%%%
%
We would call this scheme ``bidirectional control''. 
We will describe in three steps the details of the model 
which implements this scheme:\\
\hspace*{5mm} {\it Step-1: Introduction of the ``reaction'' coordinates
$\tilde{x}_{\rm L}$ and $\tilde{x}_{\rm F}$.} 
Being inspired by  Fig.~\ref{38low24.eps}, 
we introduce the convenient coordinate 
$\tilde{x}$  ($0\le \tilde{x}<\infty$) which describes
a ligand particle {\it and} a semi-detector of presence:
$x=\max(\tilde{x}-1,0),\quad$    $ a = -\min(\tilde{x},1)$. 
Then we extend this definition of $\tilde{x}$ to the case of 
two particle reservoirs:
We assume that the high-density [low-density] reservoir
occupies the half space $x>0$ [$x<0$], respectively.
We redefine the mapping $\tilde{x}\mapsto (x,a)$
 so that  $\tilde{x}$  can take the values on the entire axis, 
$-\infty<\tilde{x}<\infty$, 
with:  
\beq \label{eq:xtilde}
x=({\tilde{x}}/{|\tilde{x}|})\,\max(|\tilde{x}|-1,0), 
\quad  a = -\min(|\tilde{x}|,1).
\eeq 
A ligand particle is on the detection site if
$|\tilde{x}| \le 1$,
in the high-density reservoir if $\tilde{x} > 1$, and 
in the low-density reservoir if $\tilde{x} < -1$.
We interpret $\tilde{x}$  so that the region of 
$|\tilde{x}|\le 1-\delta$ with small $\delta >0$ 
corresponds to the state  {\sf ON} of the semi-detector, 
while $|\tilde{x}| > 1-\delta$ 
corresponds to the state {\sf OFF}. 
Finally we apply this type of mapping for both the 
{\sf load} side ($\tilde{x}_{\rm L}$) and the {\sf fuel} side
($\tilde{x}_{\rm F}$).
The coordinate plane $(\tilde{x}_{\rm L},\tilde{x}_{\rm F})$
can then represent the positioning of one representative
{\sf load} particle {\it and} the one {\sf fuel} particle 
together with the states of the semi-detectors of presence, 
$a_{\rm L}$ and $a_{\rm F}$\footnote{By the introduction of 
$\tilde{x}_{\rm L}$ and $\tilde{x}_{\rm F}$, we can
 avoid to be lost in the four-dimensional space.
The trade-off that we pay is that 
any value of $a_{\rm F} > -1$ is represented by 
the two equivalent points, 
$\tilde{x}_{\rm F}=\pm a_{\rm F}$. 
The same holds also on the {\sf load} side.}.\\
\hspace*{5mm} {\it Step-2: Definition of the gates' action.} 
For the {\sf fuel} particles, we construct the gate which allows 
the access of the particles exclusively from one of their 
reservoirs at a time.
%
%%%%%%%%%%%%%%%%%%%%%%%%%%%
\begin{figure}[ht]  
\hspace*{3.5cm}  
\resizebox{7cm}{4.75cm}{\includegraphics*{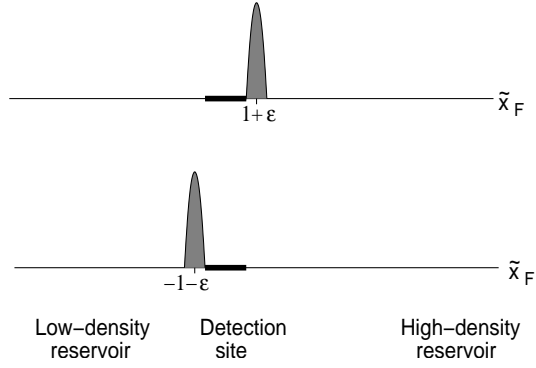}}
\caption{ {\it Top:} Potential profile of the gate that opens
to the {\sf fuel} particles in their low-density reservoir.
{\it Bottom:} Similar to the above but the accessibility is for
the high-density side.
The thick horizontal bars indicate the detection site for the 
{\sf fuel} particles. }
\label{36.eps} 
\end{figure}
%%%%%%%%%%%%%%%%%%%%%%%%%%%
%
In the {\it Top} of Fig.~\ref{36.eps}, 
the potential barrier (the hight $\gg \kT$) is established 
at $\tilde{x}_{\rm F}=1+\epsilon$ with a small $\epsilon>0$, 
so that the access of the {\sf fuel} particles 
from the high-density reservoir is blocked 
at a distance $\epsilon$ off
the detection site ($|\tilde{x}_{\rm F}|\le 1$). 
The same architecture is defined for the {\sf load} particles.
Similarly the gate represented in 
the {\it Bottom} of Fig.~\ref{36.eps} 
blocks the access of the {\sf fuel} particles from
the low-density reservoir at the distance $\epsilon$ off
the detection site.\\
\hspace*{5mm} {\it Step-3: Coupling of the semi-detectors to the gates.}  
We define the bidirectional control
by the following symmetric rules:\\
%
%%%%%%%%%%%%%%%%%%%%%%%%%%%
\begin{figure}[ht]  
\hspace*{2.5cm}  
\resizebox{9.00cm}{4.00cm}{\includegraphics*{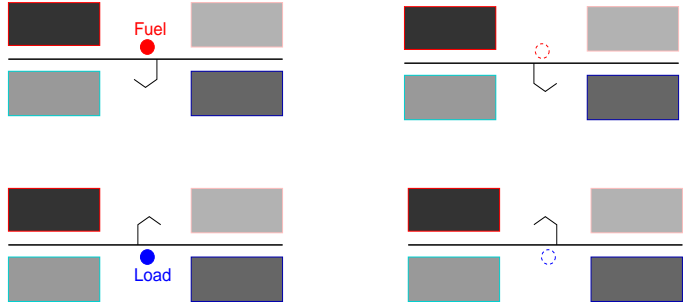}}
\caption{
{\it Top:} The control of the {\sf load} particles by the detection 
of the {\sf fuel} particles. 
{\it Bottom:} The control of the {\sf fuel} particles by the detection 
of the {\sf load} particles. 
The {\sf J}-shaped symbol in, for example,
the {\it Top-Left} indicates the exclusive accessibility 
 from the low-density reservoir of the {\sf load} particles,
as indicated in the {\it Top} of Fig.~\ref{36.eps}.
The other cases would be understood similarly.} 
\label{17.eps} 
\end{figure}
%%%%%%%%%%%%%%%%%%%%%%%%%%%
{[Control by $a_{\rm F}$]}  %
If a {\sf fuel} particle is detected, i.e. 
if $|\tilde{x}_{\rm F}| \le 1-\delta$ 
({\it Top-Left} of Fig.~\ref{17.eps}), 
 only the {\sf load} particles in the low-density reservoir 
($\tilde{x}_{\rm L}<-1$) can access to their detection site.
If a {\sf fuel} particle is not detected i.e., if 
$|\tilde{x}_{\rm F}| > 1-\delta$ ({\it Top-Right}), 
then only the {\sf load} particles in the high-density reservoir 
($\tilde{x}_{\rm L}> -1$) can access to their detection site.\\
{[Control by $a_{\rm L}$]}: %
If a {\sf load} particle is detected, i.e. if $|\tilde{x}_{\rm L}|
\le 1-\delta$ ({\it Bottom-Left} of Fig.~\ref{17.eps}), 
 only the {\sf fuel} particles in the low-density reservoir 
($\tilde{x}_{\rm F}<-1$) can access to their detection site.
If a {\sf load} particle is not detected, i.e. if 
$|\tilde{x}_{\rm L}|> 1-\delta$ ({\it Bottom-Right}), 
then only the {\sf fuel} particles in the high-density reservoir 
($\tilde{x}_{\rm F}> -1$) can access to their detection site.

\begin{figure}[ht]  
\hspace*{4.0cm}  
\resizebox{5.00cm}{5.00cm}{\includegraphics*{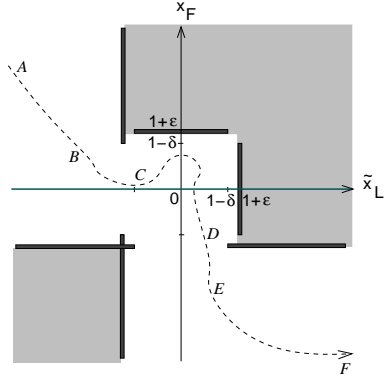}}
\caption{
In the $(\tilde{x}_{\rm L}, \tilde{x}_{\rm F})$-plane, 
the thick horizontal and vertical bars
represent the presence of the potential barriers
either for the {\sf fuel} particles (the horizontal bars) or for the 
{\sf load} particles (the vertical bars).
The shaded regions are where the bidirectional control prohibits 
to access except for a small ``leak'' due to the finite values of 
$\epsilon$ and $\delta$. See the text for the details. 
The dashed curve shows one representative process (cf. 
the footnote on the multiplication of the state points).
$B$-$E$ on the curve correspond roughly to those in 
Fig~\ref{27.eps}.
}
\label{19new.eps} 
\end{figure}

The consequence of these simple and symmetric combination 
of the semi-detectors and the gates is 
immediately seen by the graphical representation 
in Fig.~\ref{19new.eps}.
%
%%%%%%%%%%%%%%%%%%%%%%%%%%%
%%%%%%%%%%%%%%%%%%%%%%%%%%%
There, the potential barriers of the gates 
are indicated by the thick horizontal or vertical bars.
For example, the short horizontal bar at $\tilde{x}_{\rm F}=1+\epsilon$
in the range of $|\tilde{x}_{\rm L}|<1-\delta$ 
represents the blockade of the access of the {\sf fuel} particles
from their high-density reservoir.
We see how the combination of those bars
organizes a broad passageway joining the second and the forth 
quadrants on the $(\tilde{x}_{\rm L},\tilde{x}_{\rm F})$-plane.
This passageway corresponds to the diffusion
of a single {\sf fuel} particle from its high-density reservoir to the 
low-density one {\it accompanying} the active transport of a
single {\sf load} particle from its low-density reservoir to
the high-density one.
Therefore, if we could neglect the small gaps between the bars
due to the finite values of $\epsilon$ and $\delta$, this pump 
would work tightly with $\Delta N_{F,{\rm h}\to{\rm l}}=$
$\Delta N_{L,{\rm l}\to{\rm h}}$, whose mean flow direction 
is determined by the sign of 
$(\mu_{\rm F,h}-\mu_{\rm F,{\rm l}})-$ 
$(\mu_{\rm L,h}-\mu_{\rm L,{\rm l}})$.
Actual autonomous pump seems not be able to 
avoid the ``leak'' due to the above mentioned gaps,
but this leak would be of importance only near the stalled state 
($\mu_{\rm F,h}-\mu_{\rm F,{\rm l}}
=$ $\mu_{\rm L,h}-\mu_{\rm L,{\rm l}}$) 
as far as  $\epsilon$ and $\delta$ are small\footnote{In 
the context of the the linear non-equilibrium thermodynamics,
we expect that $L_{\rm sym}/L_{\rm asy}$ is at most of 
the order of $\epsilon$ and $\delta$.}.

How would the pump thus constructed 
looks like for an observer who can survey 
only the {\sf load} particles?
Let us suppose that $(\mu_{\rm F,h}-\mu_{\rm F,{\rm l}})-$ 
$(\mu_{\rm L,h}-\mu_{\rm L,{\rm l}})$ is positive and enough 
greater than \kT.
As long as the detection site of the {\sf load} particle is empty,
this site is almost always accessible from the low-density reservoir 
of the {\sf load} particle ({\sf B} along the dashed curve in
Fig.~\ref{19new.eps}).
When a {\sf load} particle arrives at the detection site 
{\it and} is detected ({\sf C,D}), the gate 
is very likely to reverse its accessibility so that 
the {\sf load} particle on the detection site can leave
now for the high-density reservoir ({\sf E}).
Then as soon as the {\sf load} particle quits the detection 
site ({\sf F}), the gate comes back to the 
initial conformation ({\sf B}).
Thus the pump behaves as if it responded by itself to the arrival
and the leave of the {\sf load} particle.

%\clp
%%%%%%%%%%%%%%%%%%%%%%%%%%%%%%%%%%%%%%%%%%%%%%%%%%%%%%%%%%%%%%%%%%%%%%%
\section{Discussion}
%%%%%%%%%%%%%%%%%%%%%%%%%%%%%%%%%%%%%%%%%%%%%%%%%%%%%%%%%%%%%%%%%%%%%%%
\label{sec:prediction}
We have constructed theoretically an autonomous system 
that works under thermal fluctuations.
The errors of the detection was avoided 
by introducing what we call the semi-detectors.
The concept of the semi-detectors on mesoscopic scale
might also be of interest in the context of 
the mesoscopic devices \citep{imry}.
The cyclic process was enabled by what we call 
the bidirectional control. 
The latter idea is applicable also to the macroscopic 
autonomous processes.
For example, in the operation protocols of vending machines
or of pay-phones, 
the gate on one side (taking up the money) and 
that on the other side (rendering goods or 
services in exchange) are controlled by
their respective detectors on the opposite sides.

The biological molecular motors are not as simple as 
we have discussed.
Still, the present analysis might help as a reference frame 
when we look for the structure-function relationship 
in those systems.
For example, we might ask if 
the hydrolysis reaction of ATP ({\sf fuel}) 
in a molecular motor corresponds to changing the accessibility 
from the reservoir of the ATP to that of 
the ADP and the inorganic phosphate [Pi].
Also we might wonder if the putative Pi-sensors 
(``switch loop'' etc., \citet{Vale1996,gilbert}) can be compared 
to the semi-detector of the ligand in our model.
As for the kinetics of the interaction between 
the molecular motor and its counterpart filament ({\sf load}), 
we might ask if the motors have
a degree of freedom to detect its own mechanical strain 
\citep{strain-detector-a,ks-prive}, 
in addition to the degree of freedom to control the 
binding to the filament \citep{KIF1A04}. 
%%%
%%%%%%%%%%%%%%%%%%%%%%%%%%%
\begin{figure}[ht]  
\hspace*{1.5cm}  
\resizebox{10.50cm}{5.55cm}{\includegraphics*{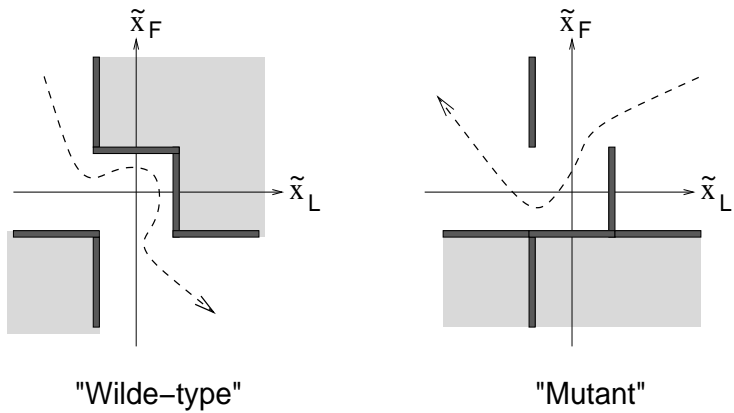}}
\caption{The pump of the ``wild-type'' 
(left, see Fig.~\protect\ref{19new.eps}) and
a ``mutant'' with blocked fuel reactions (right). See the text.} 
\label{29red.eps} 
\end{figure}
%%%%%%%%%%%%%%%%%%%%%%%%%%%
As a qualitative prediction, we might mention
the possibility of  the ``mutants'' in which
the degree of freedom related to the 
detection of, say, the {\sf load} particles, 
$a_{\rm L}$, is immobilized, as shown in Fig.~\ref{29red.eps}.
In this case, 
the rate of passive diffusion of the {\sf load} particles
should depend on the
density of the (non-consumed) {\sf fuel} particles in 
their high-density reservoirs.

I would like to thank 
K. Kitamura, H. Miyashita, E. Muneyuki, H. Noji, 
I. Ojima, Y. Okada, Y. Oono, and K. Sutoh 
for stimulating discussions.
It is my great pleasure to dedicate this paper to Y. Kuramoto.
%

%%%%%%%%%%%%%%%%%%%%%%%%%%%%%%%%%%%%%%%%%%%%%%%%%%%%%%%%%%%%%%%%%%%%%%%%%%

% Parenthetical:  \citep{Bai92} produces (Bailyn 1992).
%       Textual: \citet{Bai95} produces Bailyn et al. (1995).
         % An affix and part of a reference:  \citep[e.g.][Ch. 2]{Bar76}
         %   produces (e.g. Barnes et al. 1976, Ch. 2).

%%%%%%%%%%%%%%%%%%%%%%%%%%%%%%%%%%%%%%%%%%%%%%%%%%%%%%%%%%%%%%%%%%%%%%%%%%

%%%%%%%%%%%%%%%%%%%%%%%%%%%%%%%%%%%%%%%%%%%%%%%%%%%%%%%%%%%%%%%%%%%%%%%%%%

%\tableofcontents
\end{document}